\documentclass[sigconf]{acmart}
\usepackage[linesnumbered,ruled]{algorithm2e}
\usepackage{array}
\usepackage{CJKutf8}
\usepackage[caption=false,font=normalsize,labelfont=sf,textfont=sf]{subfig}
\usepackage{textcomp}
\usepackage[inkscapelatex=false]{svg}
\usepackage{stfloats}
\usepackage{booktabs} 
\usepackage{graphicx}
\usepackage{amsmath}
\usepackage{hyperref}
\usepackage{bbding}
\usepackage{color}
\usepackage{url}
\usepackage{verbatim}

\usepackage[markup=default,authormarkup=none]{changes}
\definechangesauthor[name={XK}, color=red]{R1}

\usepackage{multirow} 
\AtBeginDocument{%
  }
\begin{document}
\begin{CJK}{UTF8}{gbsn}
\title{An Efficient Iterative Algorithm for Qubit Mapping via Layer-Weight Assignment and Search Space Reduction}

\author{Kang Xu}
\affiliation{%
  \institution{Beijing Key Laboratory of Petroleum Data Mining, China University of Petroleum, Beijing 102249, China}
    \institution{State Key Lab of Processors, Institute of Computing Technology, CAS, Beijing 100190, China}
    \country{}
  % \city{Haidian Qu}
  % \state{Beijing Shi}
  % \country{China}
  }
  \author{Zeyang Li}
\affiliation{%
  \institution{Beijing Key Laboratory of Petroleum Data Mining, China University of Petroleum, Beijing 102249, China}
  \country{}
  % \city{Haidian Qu}
  % \state{Beijing Shi}
  % \country{China}
  }
  \author{Xinjian Liu}
\affiliation{%
  \institution{Beijing Key Laboratory of Petroleum Data Mining, China University of Petroleum, Beijing 102249, China}
  \country{}
  % \city{Haidian Qu}
  % \state{Beijing Shi}
   %\country{China}
  }
    \author{Dandan Li}
\affiliation{%
  \institution{School of Computer Science (National Pilot Software Engineering School)}
    \institution{Beijing University of Posts and Telecommunications, Beijing 100876, China．}
    \country{}
  %\country{China}
  }
     \authornotemark[1]
     \email{dandl@bupt.edu.cn}
  \author{Yukun Wang}
\affiliation{%
  \institution{Beijing Key Laboratory of Petroleum Data Mining, China University of Petroleum, Beijing 102249, China}
  \institution{State Key Lab of Processors, Institute of Computing Technology, CAS, Beijing 100190, China}
  \country{}
  % \city{Haidian Qu}
  % \state{Beijing Shi}
  % \country{China}
  }
   \authornotemark[2]
     \email{wykun06@gmail.com}

\begin{abstract}

{Current quantum devices support interactions only between physically adjacent qubits, preventing quantum circuits from being directly executed on these devices. Therefore, SWAP gates are required to remap logical qubits to physical qubits, which in turn increases both quantum resource consumption and error rates. To minimize the insertion of additional SWAP gates, we propose HAIL, an efficient iterative qubit mapping algorithm. Leveraging the inherent parallelism in quantum circuits, a new layer-weight assignment method is integrated with subgraph isomorphism to derive an optimal initial qubit mapping. Moreover, we present a two-stage SWAP sequence search algorithm that effectively identifies the most efficient SWAP sequence by distilling feasible SWAP sequences at different stages. The whole qubit mapping algorithm is then refined through a few iterative bidirectional traversals, further reducing the number of SWAP gates required. Experimental results on the IBM Q20 architecture and various benchmarks show that HAIL-3 reduces the number of additional gates inserted in the $\mathcal{B}_{23}$ by 20.62\% compared to state-of-the-art algorithms. Moreover, we propose a partially extended SWAP sequence strategy combined with HAIL to reduce its time complexity, with experiments on the sparsely connected Google Sycamore architecture demonstrating reductions in both algorithm runtime and additional SWAP gates.}

\end{abstract}


\pagestyle{empty}
\keywords{Qubit mapping, subgraph isomorphism, iterative optimization, heuristic algorithm.}


\settopmatter{printacmref=false} 
\renewcommand\footnotetextcopyrightpermission[1]{}

\maketitle
\section{Introduction}
In recent years, quantum computing has emerged as a leading post-Moore's Law technology, attracting significant attention from researchers due to its powerful parallel processing capabilities. The use of quantum computers to simulate quantum systems, initially proposed by Feynman \cite{feynman2018simulating} in 1982, has driven significant advancements in the field. Examples of such progress include Shor’s algorithm for factoring \cite{shor1994algorithms}, Grover’s search algorithm \cite{grover1996fast}, and Harrow’s quantum linear systems algorithm \cite{harrow2009quantum}. In addition to the development of quantum algorithms \cite{li2025efficient,song2024quantum}, quantum hardware has also undergone significant improvements.
Recent milestones include the development of quantum chips with {70-100 }
qubits across diverse quantum technologies such as superconducting, neutral atoms, ion traps, and photonics. This progress culminated in 2023 with the launch of IBM's Condor \cite{Condor2023ibm} quantum processor, featuring 1121 qubits and significantly advancing the practical application of quantum computing.

Despite the remarkable progress in quantum computing, achieving large-scale fault-tolerant quantum computing still faces significant challenges such as limited qubit count and inadequate device connectivity. When executing quantum algorithms on real quantum devices, the connectivity of the physical qubits must be considered, as prevailing quantum architectures support multi-qubit gates only between interconnected qubits. Therefore, it is essential to first establish a mapping between the logical qubits and the physical qubits of the quantum device. When a multi-qubit gate does not satisfy the current mapping relationship, SWAP gates must be inserted to adjust the mapping relationship and make non-adjacent qubits adjacent, enabling the execution of circuits while maintaining functional equivalence between physical and logical circuits. However, this qubit mapping process introduces additional gates, increasing both gate count and circuit depth \cite{sun2023asymptotically}, which can adversely affect quantum algorithm performance. So the primary goal of qubit mapping is to minimize the number of SWAP gates.

The qubit mapping problem, formally recognized as NP-complete \cite{botea2018complexity, NPsiraichi2018qubit}, poses significant challenges. Various methods \cite{ge2024quantumcircuitsynthesiscompilation} have been employed to address this issue, including mathematical programming, machine learning, and heuristic algorithms.
Mathematical programming techniques involve converting mapping problems into integer programming problems \cite{shafaei2014qubit,nannicini2022optimal,bhattacharjee2017depth, bhattacharjee2019muqut}, satisfiability (SAT) problems \cite{tan2021optimal,murali2019noise,lye2015determining,wille2019mapping,molavi2022qubit,lin2023scalable}, and then solving them with different kinds of advanced solver, such as the SMT solver (Satisfiability Modulo Theories) and the SAT solver. However, these methods are often limited to smaller problems due to their high time complexity.
%However, these methods are typically used for problems with a few qubits and quantum gates because of their high time complexity in handling mapping problems. 
On the other hand, machine learning approaches for qubit mapping problems have emerged with the advancement of artificial intelligence technologies \cite{zhou2020monte,sinha2022qubit,pozzi2022using,li2024deep,fan2022optimizing,zhou2022supervised,paler2023machine}, including Monte Carlo Tree Search \cite{zhou2020monte}, reinforcement learning \cite{pozzi2022using}, and neural networks \cite{zhou2022supervised}, for decision making. However, these methods face challenges such as extensive data requirements, high computational costs, and limited interpretability, which hinder their widespread adoption. %machine learning approaches for qubit mapping problems have emerged with the advancement of artificial intelligence technologies\cite{zhou2020monte,sinha2022qubit,pozzi2022using,li2024deep,fan2022optimizing,zhou2022supervised,paler2023machine}, encompassing Monte Carlo Tree Search \cite{zhou2020monte,sinha2022qubit}, reinforcement learning techniques \cite{pozzi2022using,li2024deep,fan2022optimizing}, and artificial neural networks\cite{zhou2022supervised} for decision making.
%However, machine learning methods also face challenges such as extensive data requirements for model training, substantial computational resources, and a lack of intuitive interpretability, which prevents them from becoming the predominant research approach.

In contrast, heuristic algorithms offer excellent scalability, efficiently solving mapping problems involving tens of qubits and thousands of quantum gates in a relatively short time. 
These algorithms use experience-based rules to optimize qubit placement and SWAP gate insertion for near-optimal SWAP costs. Numerous heuristic approaches have been proposed \cite{li2019tackling,li2020qubit,qian2023method,zulehner2018efficient,zhu2021iterated,zhu2020dynamic,matsuo2012changing,ash2019qure,niu2020hardware,lao20222qan,li2023timing,jiang2021quantum,datta2023improved,chang2021mapping,ovide2023mapping,liu2023tackling,huang2024efficient}, {most of which decompose the problem into two phases: initial mapping and qubit routing. %They typically decompose the problem into two phases: initial mapping, which assigns logical qubits to physical qubits, and qubit routing, which inserts SWAP gates to resolve connectivity constraints—all guided by experience-based heuristics to approach near-optimal SWAP costs.
%Numerous heuristic algorithms \cite{li2019tackling,li2020qubit,qian2023method,zulehner2018efficient,zhu2021iterated,zhu2020dynamic,matsuo2012changing,ash2019qure,niu2020hardware,lao20222qan,li2023timing,jiang2021quantum,datta2023improved,chang2021mapping,ovide2023mapping,liu2023tackling,huang2024efficient} for qubit mapping have been developed.
In the initial mapping, logical qubits are assigned to physical qubits such that the early portion of the circuit can be directly executed. For example, SABRE \cite{li2019tackling} starts with a random mapping and refines it iteratively through a bidirectional search that considers gate priority and hardware topology. Alternatively, subgraph isomorphism techniques \cite{li2020qubit} generate mappings by aligning the circuit’s qubit interaction graph with the device’s coupling graph.  However, due to physical connectivity constraints, the initial mapping is rarely sufficient, necessitating a qubit routing phase where SWAP gates are inserted to enable the execution of subsequent gates. Various techniques exist for SWAP insertion. The A* search-based method \cite{zulehner2018efficient} yields low-cost SWAP placements but suffers from an exponentially growing search space. To mitigate this issue, many studies restrict each routing step to a single SWAP gate, thereby reducing the candidate search space; however, this often leads to a higher overall SWAP cost. Thus, filtered depth-limited heuristic searches have been proposed, where some depth-limited combinations of SWAP gates are considered. Taking \cite{li2020qubit,qian2023method} for example, which explores the simultaneous insertion of multiple SWAPs involves some nearest layers to reduce the total SWAP count. Iterative algorithms that integrate bidirectional search with shuffling perturbations \cite{zhu2021iterated} further minimize SWAP costs but typically require high iteration counts and extended runtimes. Despite these advances, several challenges persist, including neglect of parallel gate weights, difficulty in determining appropriate search space sizes, and performance bottlenecks observed in benchmarks such as QUEKO \cite{tan2020optimality}. These limitations highlight the need for continued refinement of heuristic qubit mapping strategies to improve mapping quality, scalability, and computational efficiency.}

In this paper, we propose HAIL, an iterative optimization-based heuristic algorithm with the aim of minimizing the additional SWAP gates in the qubit mapping problem. The algorithm begins by introducing a subgraph isomorphism algorithm and a completion algorithm where qubit insertion around placed qubits, with layer-weight assignment to partition the circuit, and map it into a coupling structure graph. %To address limitations of the subgraph isomorphism algorithm, a mapping completion approach is employed to enable qubit insertion around placed qubits. 
Once a complete initial mapping is achieved, HAIL partially expands the SWAP sequence space to shrink the candidate set, then conducts a sequence search that prioritizes sequences with the highest average executable CNOT count, and refines this selection with a scoring function based on distance and layer-weight.  Finally, a simplified iterative framework with forward-backward traversal is applied over a few iterations to identify the configuration that minimizes the overall SWAP gate cost. Experimental results show that HAIL-3 outperforms state-of-the-art methods, including SABRE, ILS, and TWP, achieving a 20.62\% reduction in additional SWAP gates on the IBM Q20 for the $\mathcal{B}_{23}$ benchmark. 

The rest of this paper is organized as follows. Section 2 introduces the background of quantum computing, followed by a detailed description of the proposed algorithm in Section 3. In Section 4, experiments are carried out on the IBM Q20 and Google Sycamore architectures, with a comprehensive analysis of the results performed. Finally, the whole paper is summarized.
\begin{figure}[!t]
\centering
\includegraphics[width=\linewidth]{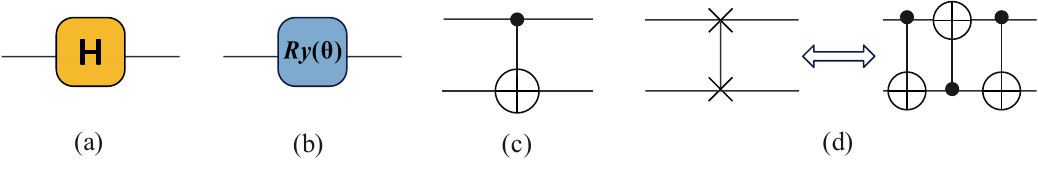}
\caption{Some common basic gates. (a) H gate, (b) Ry($\theta$) gate, (c) CNOT gate, (d) SWAP gate.}
\label{fig_1}
\end{figure}
\begin{figure}[t]
\centering
\includegraphics[width=0.95\linewidth]{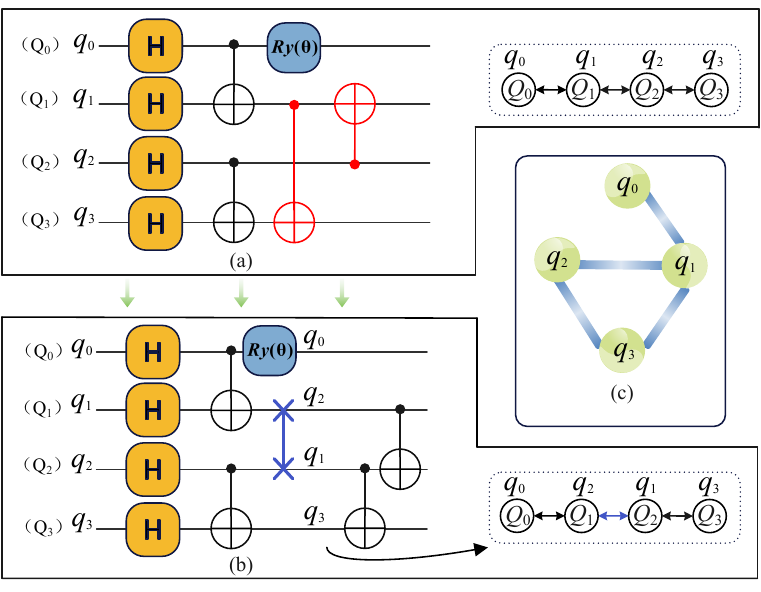}
\caption{(a) An initial circuit and one of its corresponding mappings to the linear quantum architecture, where the red color indicates that the gates are not executable under the mapping in this architecture,
(b)  A swap is inserted to enable execution in the linear quantum architecture and mapped accordingly, (c) The qubit interaction graph of the initial circuit.}
\label{fig_2}
\end{figure}
\begin{figure}[t]
\centering
\includegraphics[width=\linewidth]{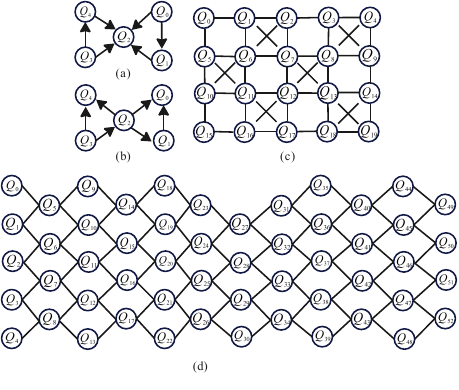}
\caption{The quantum hardware architectures. (a) IBM QX2, (b) IBM QX4, (c) IBM Q20, (d) Google Sycamore.}
\label{fig_4}
\end{figure}
\section{Background}
This section aims to provide a concise introduction to the fundamentals of quantum computing, enabling a better understanding of the challenges surrounding qubit mapping and the proposed methodologies.
% to aid in understanding the issues of qubit mapping and the subsequent proposed algorithms.
\subsection{{Quantum Gates and Quantum Circuits}%
}
Qubit is the fundamental unit of quantum computing. Unlike classical bits, which can only be in states of 0 or 1, qubits can exist in superpositions of these states due to quantum mechanics. In quantum computing, a qubit's state can be represented as $|\phi\rangle = \alpha|0\rangle + \beta|1\rangle$, where $|0\rangle$ and $|1\rangle$ are the basic states, and $\alpha$ and $\beta$ are complex numbers satisfying the normalization condition $|\alpha|^2 + |\beta|^2 = 1$.
% Similarly, when extending this concept to the scenario involving two qubits, the system's state can be described as $|\phi_2\rangle = \alpha_{00} |00\rangle + \alpha_{01} |01\rangle +\alpha_{10} |10\rangle + \alpha_{11} |11\rangle$. Moreover, it remains subject to $|\alpha_{00}|^2 + |\alpha_{01}|^2 + |\alpha_{10}|^2 + |\alpha_{11}|^2 = 1$.

Quantum gates typically act on qubits, and a combination of single-qubit gates and CNOT gates can generally achieve arbitrary unitary gates. Therefore, these two types of quantum gates are introduced primarily.
% \begin{figure}[!t]
% \centering
% \includegraphics[width=0.50\linewidth]{try/Fig/SWAP.png}
% \caption{The SWAP gate can be decomposed into three CNOT gates.}
% \label{SWAP}
% \end{figure}
In Figure \ref{fig_1}, some representative quantum gates are illustrated. The $H$ gate is extensively used for creating quantum superposition states. The $Ry(\theta)$ gate, a single-qubit gate rotating around the Y-axis, is prominent in variational quantum algorithms. The CNOT gate, denoted as $g=<p,q>$, alters the target qubit's state based on the control qubit's state. If the control qubit is $|1\rangle$, the target qubit undergoes an $X (NOT)$ gate. The SWAP gate facilitates the exchange of states between two qubits, transforming $|a, b\rangle$ to $|b, a\rangle$. Additionally, a SWAP gate can be decomposed into three CNOT gates, which is crucial in qubit mapping problems.

A quantum circuit acts as a formal representation of a quantum algorithm or program, delineated by a series of quantum gates, as demonstrated in Figure \ref{fig_2}(a). This circuit comprises five single-qubit gates ($H$ gate and $Ry(\theta)$ gate) and four two-qubit gates (CNOT gate). In this paper, such quantum circuits are called logical circuits, denoted by the notation $LC(Q, G)$, where $Q$ represents logical qubits, and $G$ denotes the set of all gates in the circuit. Figure \ref{fig_2}(b) shows the quantum circuit after being adjusted for execution in the given quantum architecture, where a SWAP gate is inserted. Since gate permutation or cancellation operations between gates are not performed in the quantum circuit in our paper, single-qubit gates do not affect the outcome of qubit mapping. Thus, the qubit mapping problem can be simplified to only consider two-qubit gates.
\subsection{Qubit Interaction Graph and Quantum Hardware Architecture}
Once a quantum logical circuit $LC(Q, G)$ is obtained, a qubit Interaction Graph ($IG$) can be defined to represent the connections of the logical qubits. The \(IG\) is an undirected graph where nodes represent the qubits in \(Q\). An edge is formed between two nodes \(q_i\) and \(q_j\) if there is a correlated two-qubit gate between them, which plays a crucial role in subgraph isomorphism matching. Specifically, Figure \ref{fig_2}(c) illustrates the \(IG\) for the quantum circuit shown in Figure \ref{fig_2}(a), consisting of four nodes and four edges.

{The quantum hardware architecture is represented as $AG(V, E)$, where $V$ denotes the nodes and $E$ denotes the edges.} To provide an overview of quantum hardware architectures, Figure \ref{fig_4} illustrates several architectures from IBM and Google. Despite advancements such as the Condor quantum processor, current architectures are still constrained by connectivity limitations. Among these architectures, the IBM Q20 is notable for its dense bidirectional connectivity and is widely used.

\subsection{Initial Qubit 
 Mapping and Routing Problem}
The quantum circuit mapping problem is primarily divided into the initial qubit mapping and qubit routing problems, with the assumption that the input quantum circuit is denoted as \( LC(Q, G) \) and the given quantum device as \( AG(V, E) \).

First, the initial qubit mapping problem refers to the one-to-one correspondence between logical qubits and physical qubits (e.g., \( Q \rightarrow V \)). This process typically involves various approaches such as random mapping, naive mapping, and optimization-based techniques. An example of a simple mapping is shown in Figure \ref{fig_2}(a), where the mapping is represented as \( \tau = \{q_0 : v_0, q_1 : v_1, q_2 : v_2, q_3 : v_3\} \). Under this mapping, the gate marked in red cannot be executed because the mapped nodes are not directly connected.

The routing problem is defined as follows: during the execution of the circuit, if a gate cannot be executed due to the current mapping and gate dependency constraints, SWAP gates must be inserted to transform the mapping, resulting in a new mapping \( \tau' \). As shown in Figure \ref{fig_2}(b), the insertion of SWAP gates allows the remaining two CNOT gates to be executed. However, this process introduces additional SWAP gate operations, and thus, the primary optimization goal of quantum circuit mapping is to minimize the number of additional SWAP gates inserted.

\section{Method}
This section provides an introduction to the proposed method for minimizing the additional insertion of SWAP gates. First, a layer-weighted subgraph isomorphism approach and a novel mapping completion method are presented to establish the initial mapping. Subsequently, a SWAP sequence search method is introduced to address the selection of SWAP gates. Finally, the approach incorporates a simplified iterative optimization framework, which significantly enhances the overall effectiveness of the mapping algorithm.

\begin{figure*}[t]
\centering
\includegraphics[width=0.95\linewidth]{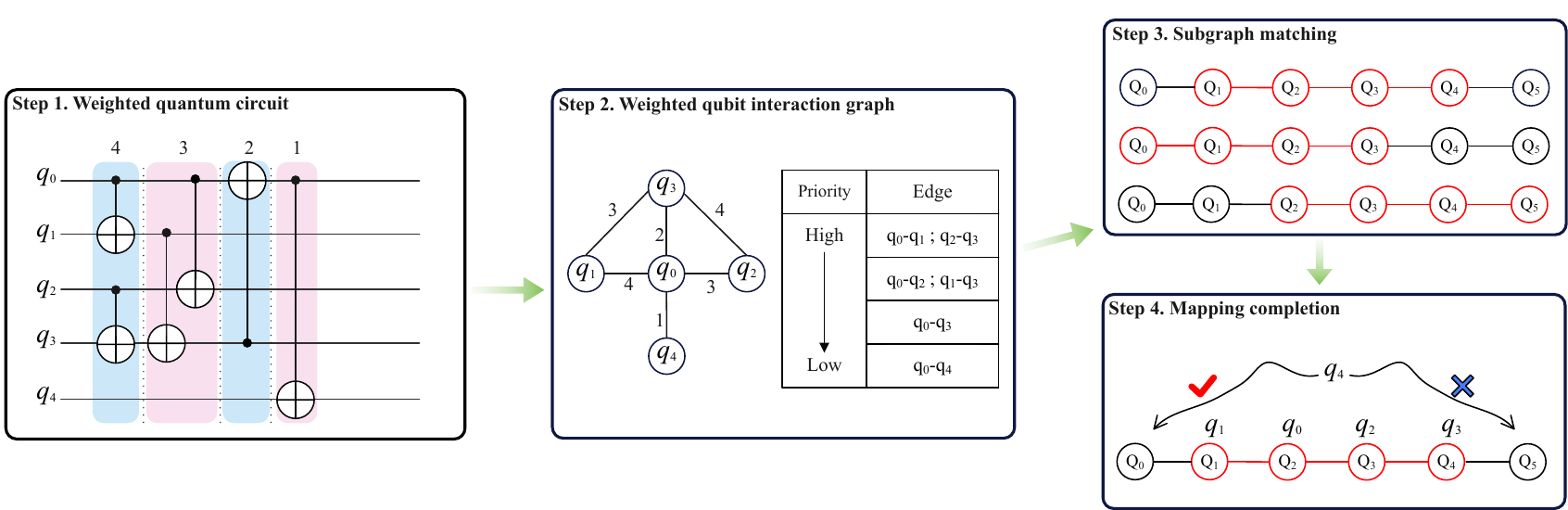}
\caption{Paradigm of initial mapping generation. In Step 1, each gate in the quantum circuit is prioritized based on layer-weight assignment, with the number on the gate representing its weight. In Step 2, for the qubit interaction graph generated by the logical circuit, the weight on each edge is the sum of the weights of the gates acting on the respective qubits. In Step 3, the matching of the subgraph for a specific architecture is completed by gradually adding edges with higher priorities. In Step 4, for the unmapped qubits, their placement locations are determined to complete the mapping.}
\label{fig_6}
\end{figure*}
\subsection{Initial Mapping}
To address the mapping problem, we divided the entire algorithm into two parts, ultimately generating a reasonable initial mapping to reduce the number of subsequent SWAP gate insertions.
\subsubsection{\bf{Layer-Weighted Isomorphism Mapping Algorithm}}
We are addressing the connectivity issue and have proposed a layer-weight-based method to fully consider the priority of each gate in the circuit through weight assignment, %\
{thereby enabling the execution of more CNOT gates at the front of the circuit and reducing the distance between subsequent gates.}
%N
This method prioritizes quantum gates located at the front of the circuit, ensuring that they have the highest priority. When multiple gates can be executed simultaneously, they are assigned equal weights, making this weight-based layer approach intuitive and rational. The process is detailed in Steps 1 to 3 in Figure \ref{fig_6}. Initially, the circuit is partitioned, ensuring qubits of any two gates in the same layer do not intersect. Then, for each gate $g_i$, a corresponding weight is assigned using the following weight function:
\begin{equation}
\label{eq1}
W_{g}(g_i )=(Depth(LC)-Layer(g_i ))+1,
\end{equation}
where $Depth(LC)$ represents the circuit depth after the partition of $LC$, and $Layer(g_i)$ represents the layer number where $g_i$ is located.

Then for any given quantum circuit $LC$, after denoting $IG(Q, E)$ to describe the interactions between qubits in the quantum circuit, we use the layer weight (Eq. \ref{eq1}) to assign a value for each gate in the circuit. In $IG$, the weight on each edge is the sum of the weights of the corresponding gates, described by the following formula:
\begin{equation}
\label{eq2}
W_{e}((p,q))=\sum_{p\, \&\, q\ in\ g_i}^{LC} {W_{g}(g_i)},
% E_{weight}((p,q))=\sum_{(p,q)\ in\  g_i}{G_{weight}(g_i)},
\end{equation}
where the criterion implies that both qubits $p$ and $q$ should be involved in the quantum gate $g_i$, and $(p,q)$ represents the edge between the two qubits. 
% Figure \ref{fig_7} clearly illustrates the layer-weight-based assignment method. By evaluating the priority of each gate in the circuit, the weight of each edge in $IG$ is obtained.

% After obtaining the $IG$ with layer-weight assignment, the subgraph isomorphic matching algorithm is introduced to accomplish the task of initial mapping generation.
% Subgraph isomorphism matching aims to find a subgraph in a larger graph $G_{big}$ that exactly matches the structure of a given smaller graph $G_{small}$. Generally, if $G_{small}$ can be embedded in $G_{big}$, then there exists a subgraph in $G_{big}$ that is isomorphic to $G_{small}$. 
% The commonly used subgraph isomorphic matching algorithm is the VF2\cite{1323804} algorithm. 
%Although the subgraph isomorphism matching problem is NP-complete, algorithms such as VF2 can effectively tackle graph matching problems with thousands of nodes within a reasonable timeframe.  Given that current quantum computing development is in the NISQ era, where most quantum circuits involve fewer than 100 qubits, the VF2 algorithm can swiftly and effectively accomplish the task.
% , so it is wise to adopt this algorithm.
\begin{algorithm}[t]
    \SetAlgoLined %显示end
	\caption{Initial mapping employing subgraph isomorphism and layer-weight }%算法名字
        \label{alg1}
	\KwIn{A logic circuit $LC$, a architecture graph $AG$.}%输入参数
	\KwOut{A mapping of logical qubits to physical qubits $result$.}%输出
$LC\_{depth}, Layer = Partition(LC)$\;
$W_e = \{\,\}, G_{im} = \{\,\}$\;
$IG = LC\_to\_IG(LC)$\;
 %\;用于换行
	\For{$g$ in LC}{
$W_g(g) = LC\_depth - Layer(g) + 1$\;
		\eIf{ $(g_{\cdot} c, g_{\cdot} t )$ in IG }{
		 $W_e((g.c, g.t)) += W_g(g)$\;
		}{
 $W_e((g.t,g.c)) += {W_g(g)}$;}
 }
$sort(W_e)$\;
\For{$edge\ in\ W_e$}{
\If{ $subgraph\_isomorphism (G_{im} \cup edge,\ AG)$ }{$G_{im} = G_{im} \cup edge$\;}
}
{$temp\_result = vf2_{\cdot} dfsMatch(G_{im},\ AG)$}\;
{$result = Finalizing(AG,\ IG,\ W_e,\ temp\_result)$}\; 
{\textbf{return} $result$}\;
\end{algorithm}
After obtaining the $IG$ with layer-weight assignment, we utilize the VF2 algorithm \cite{1323804}, a widely used subgraph isomorphic matching algorithm, to maximize the embedding of edges contained in the $IG$ into the $AG$, thus establishing the correspondence between logical qubits and physical qubits. The initial mapping algorithm is described in Algorithm \ref{alg1}. It takes the logical circuit and a specific hardware architecture as inputs. Initially, the logical circuit is partitioned into layers using the function $Partition$, which determines the depth of the quantum circuit and assigns each gate to its corresponding layer number. Subsequently, the function $LC\_ to\_ IG$ converts the logical circuit into a qubit interaction graph ($IG$). Each gate is then assigned a weight based on its layer position, and these weights are associated with the edges in $IG$. Finally, the edges in $IG$ are sorted based on their weights in descending order, establishing the priority of each edge. Upon successive addition of each edge to $W_e$ to form the graph $G_{im}$, which represents the current maximum subgraph matched. If the addition of an edge makes $G_{im}$ isomorphic to any subgraph in the hardware architecture $AG$, then the edge is included in $G_{im}$. This process continues until all edges have been traversed. 

{{For example, }%
in step 3 of Figure \ref{fig_6}, we begin by incorporating the edge set $\{(q_0, q_1), (q_2, q_3)\}$ to form the graph $G_{im}$, based on the priority of edges obtained in step 2, because these two edges can be embedded into the architecture graph. $(q_0, q_2)$ is then added to {$G_{im}$}%
, and it remains compatible with the architecture graph. For the subsequent edges $(q_1, q_3)$, $(q_0, q_3)$, and $(q_0, q_4)$, adding any of these to $G_{im}$ does not fit within the architecture graph, so these edges are excluded from $G_{im}$. Ultimately, a graph is obtained that includes the edges $\{(q_0, q_1), (q_0, q_2), (q_2, q_3)\}$, which can be embedded in \(AG\), while a qubit labeled \(q_4\) remains unmapped. The red parts illustrate the feasible mapping relationships, one of which is represented by the mapping $\tau = \{q_1: v_1, q_0: v_2, q_2: v_3, q_3: v_4\}$. This mapping enables the execution of three CNOT gates, whereas the naive mapping can only execute two, thus making the proposed approach more efficient.}
\begin{algorithm}[t]
    \SetAlgoLined %显示end
	\caption{Finalizing the initial mapping}%算法名字
        \label{alg2}
	\KwIn{A architecture Graph $AG$, a interaction Graph $IG$, an edge weighted graph E$_w$, a transient mapping T$_m$. }%输入参数
	\KwOut{A modified final mapping $result$. }%输出
{$result = T_m$}\;
\While{$existing\ unmapped\ qubits$}{
{$maximum = 0$}\;
$free\_place = adjacent(result,\ IG)$\;
\For{$qubit\ in\ IG.nodes$}{
$T = result$\;
\If{ $qubit\ is\ not\ mapped$}{
\For{$node\ in\ free\_place$}{
$T.update({qubit: node})$\;
$F = F_{val}(T,\ E_w,\ AG)$\;
\If{$F\,\textgreater\,maximum$}{
$maximum = F$\;
$cand = \{qubit: node\}$;
}
}
}
}
$result.update(cand)$\;
}
{
\textbf{return} 	$result$\;
}
\end{algorithm}
\subsubsection{\bf{Mapping Completion Algorithm}}
Since subgraph isomorphism may result in discrete unmapped qubits during the initial mapping process, we introduce a completion algorithm to address this issue.
{The primary objective is to map the unmapped qubits to the most suitable free nodes, thereby reducing the distances between gates involving these qubits and subsequently minimizing the number of SWAP gates. As the two most critical attributes of gates are distance and weight, the optimal scenario for circuit execution is to enhance the number of executable gates by reducing the distances for gates with greater weights.}
In Step 4 of Figure \ref{fig_6}, the mapping completion algorithm serves a discriminative function to determine the optimal positions. The algorithm details are outlined in Algorithm \ref{alg2}.

{Considering that a closer placement of qubits leads to a reduction in the distance between the two qubits of the CNOT gates}, the candidate nodes refer to the free nodes that are located adjacent to the already deployed nodes. For all logical qubits, if the qubit has not been mapped yet, the corresponding function value is calculated for placing it on each available node, and the position \(v\) with the maximum function value is identified for \(q\). The detailed content of the function $F_{val}$ is as follows:
% argmax_{v} 
\begin{equation}
\label{eq3}
F_{val}(q,v)=\sum_u(dia(AG)-dis(u,v))*E_w(u,v), 
\end{equation}
where $q$ represents an unmapped qubit, $u$ is the node in $AG$ that has been mapped by $q$'s neighboring logical qubit, and $v$ is a previously mentioned free candidate node in $AG$. The $dia(AG)$ stands for the diameter of $AG$, $dis(u,v)$ is the distance between the nodes $u$ and $v$, and $E_w(u,v)$ represents the edge weight between logic qubits corresponding to the nodes $u$ and $v$.

The proposed scoring function effectively accounts for the relationship between the evaluation node positions and their corresponding weights. If the distance between the free node $v$ and the already placed node $u$ is short, and the weight value between them is high, it indicates that the mapping can execute more CNOT gates. Therefore, selecting the logical qubit $q$ and the free node $v$ with the highest function value enables a good completion of the current mapping algorithm, and adding this mapping relation to the current mapping completes the update. This process continues iteratively until all logical qubits are mapped to appropriate free nodes. Ultimately, the algorithm returns a final mapping result, where each logical qubit is mapped to a corresponding physical hardware node, completing the mapping process.

{For example, as illustrated in step 4 of Figure \ref{fig_6}, the placement position for the unallocated qubit $q_4$ is determined using the evaluation function specified in Eq. \ref{eq3}. 
The function is calculated for two potential mapping positions:
\begin{align}
F_{val}(q_4 \rightarrow v_0) = (5 - 2) \times 1 = 3.\\
F_{val}(q_4 \rightarrow v_5) = (5 - 3) \times 1 = 2.
\end{align}
These calculations reveal that mapping qubit $q_4$ to $v_0$ results in a higher evaluation score of 3, in contrast to a score of 2 for mapping it to $v_5$. To optimize the evaluation function and ensure the most effective placement, $v_0$ is chosen as the mapping position for qubit $q_4$. As a result, the final mapping relationship is expressed as $
\tau = \{q_1: v_1, q_0: v_2, q_2: v_3, q_3: v_4, q_4: v_0\},
$ which demonstrates that it is a complete mapping and each logical qubit is assigned a specific position within the architecture graph.}

\subsection{Intermediate Qubit Routing}
The initial mapping obtained above typically fails to meet the requirements for executing all CNOT gates, requiring additional SWAP gates.  {Unlike previous approaches that rely solely on a single scoring function to determine SWAP gates, our proposed qubit routing algorithm introduces a progressively refined selection process to minimize the total number of inserted SWAP gates. The algorithm consists of two stages: SWAP sequence selection and post-processing to refine the optimal sequence.}
%\
% ensured to be 
\begin{figure}[!t]
\centering
\includegraphics[width=\linewidth]{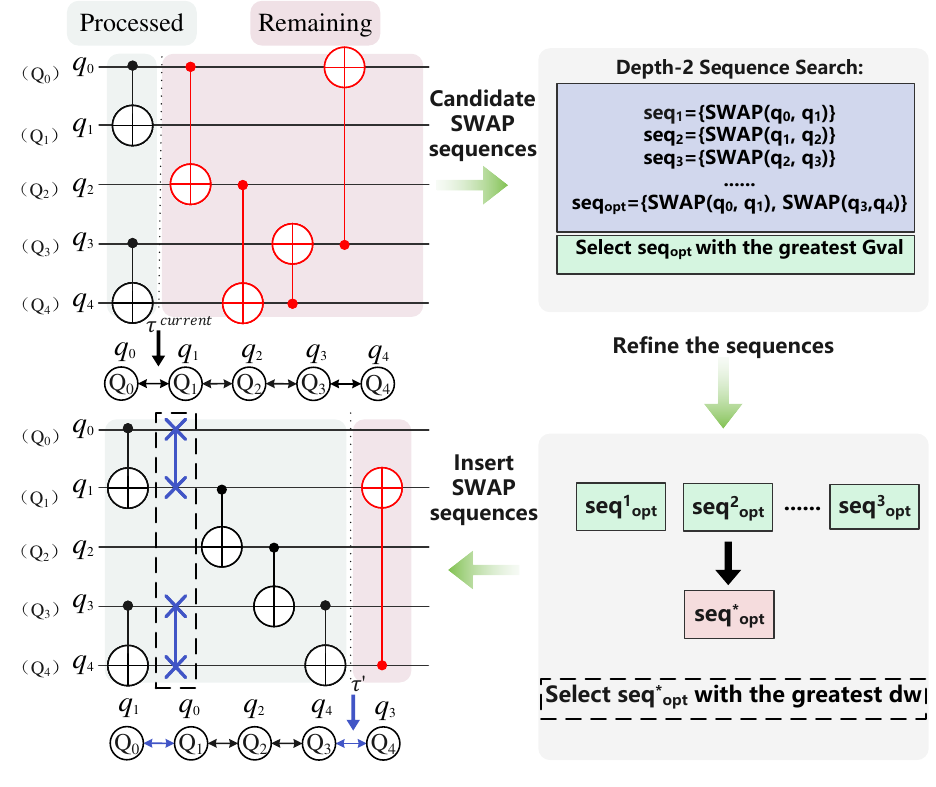}
\caption{{Diagram of SWAP sequence insertion. Four CNOT gates cannot be executed in the circuit. After determining the optimal SWAP sequence through a Depth-2 sequence search and the subsequent post-processing stage, the mapping relationship is adjusted to ensure the execution of three CNOT gates.
%y
}%
}

\label{fig_8}
\end{figure}

\subsubsection{\bf SWAP Sequence Selection}
We introduce a method for selecting SWAP sequences to better accommodate the initial mapping algorithm proposed in the previous section. First, we introduce the set \( Edges_{tp} \)where each element represents a SWAP gate that involves at least one qubit from the look-ahead three layers of the currently unexecutable circuit. We also assume that the number of elements in this \( Edges_{tp} \) is $t$. Therefore, for a SWAP sequence of length $d$, it is formulated as $seq_i = \{ \text{SWAP}_1, \text{SWAP}_2, \ldots, \text{SWAP}_d \} $, with each $\text{SWAP}_k \in  Edges_{tp} $. There are a total of $O(t^d)$ possible $d-$length swap sequence, and the collection of all such sequences forms $SQ_d$. For the SWAP sequence search with depth $d$, the set is defined as $seq = \{ SQ_1, SQ_2, \ldots, SQ_{d} \}$, where the number of elements in $seq$ is $O(t+t^2+\ldots+t^d)$. For a quantum device $AG(V, E)$, the total number of candidate SWAP sequences is $O(|E|^d)$ in the worst case.%and the definition of Depth-$n$ SWAP sequence search is as follows: 

For the evaluation of each SWAP sequence, the average number of two-qubit gates executable per SWAP gate is selected as the primary metric. This approach is considered more direct and insightful than a distance-based evaluation between gates. Consequently, the heuristic function is defined as follows, with the optimal sequence $seq_{opt}$ chosen to maximize this heuristic function:
\begin{equation}
\label{eq4}
% Gval(\tau^{current},seq)=\frac{num\ of\ gates\ executable\ by\ \tau^{update}}{len (seq) }.which may not satisfy subsequent gate requirements, necessitating transformations to enable their execution
Gval(\tau^{current},seq_i)=\frac{{\rm Executable\ gate\ count\ under}\ \tau^{'}}{len (seq_i) },
\end{equation}
where $\tau^{current}$ denotes the current mapping, $len(seq_i)$ indicates the number of SWAP gates contained in the $seq_i$. \( \tau' \) represents the mapping relation corresponding to \( \tau^{current} \) after the insertion of the SWAP gates included in \( seq_i \).

Finally, after determining the SWAP sequence $seq_{opt}$ corresponding to the maximum function, the SWAP gates in the $seq_{opt}$ are inserted sequentially. However, since the calculation may have multiple maximum function values $Gval$, further judgments are required to determine which SWAP sequence is optimal. This additional processing is conducted in the post-processing stage to make a more reasonable decision. Besides, if a suitable SWAP sequence cannot be found within the set search limit to enable the subsequent gates to be executed, an alternative approach is adopted. {The CNOT gate whose corresponding two qubits are closest }%啥意思
%NOTE：也就是一个CNOT门对应两个结点，然后两个结点会有距离。
in the front layer is selected, and then a SWAP gate is applied {to reduce the distance between them.}% 
%N
This ensures that the routing algorithm runs smoothly, preventing situations where CNOT gates cannot be executed within the specified search depth.
\begin{algorithm}[t]
  \SetAlgoLined
  \caption{Routing operation optimization}
  \label{alg3}
  \KwIn{A logical circuit $LC$, an architecture graph $AG$, an initial mapping of input $\tau^{initial}$. }
  \KwOut{A physical circuit adapted to a specific architecture $PC$, the mapping at the end of the quantum circuit $\tau^{current}$.}
$\tau^{current} = \tau^{initial}$\;
$PC = [\,]$\;
  \While{$existing\ unexecuted\ gates$}{
    $EG = Executable\_gates(LC,\ AG,\ \tau^{current})$\;
    \For{$gate\ in\ EG$}{$LC.delete(gate)$\;
				$PC.add(gate)$\;
}
    \If{$len(LC)\, ==\, 0$}{break\;}
    Find the maximum swap sequence $seq_{opt}$ that satisfies the heuristic function Eq. \ref{eq4}\;
    $SP = argmax\ dw(LC,\ AG,\ \tau^{current},\ seq_{opt})$\;
    $PC.add(SP)$\;
   $\tau^{current}=transformation(\tau^{current},\ SP)$\;

    }
\textbf{return} $PC,\ \tau^{current}$\;

\end{algorithm}
\subsubsection{\bf The Post-Processing Stage} 
 {The purpose of this stage is to distill the mapping relationships corresponding to SWAP sequences with the same maximum Gval value from the previous SWAP sequence selection stage. However, randomly selecting one of these SWAP sequences as an insertion sequence may result in an increased distance between quantum gates at the rear part of the circuit, thereby increasing the number of additional SWAP gates and potentially degrading the algorithm's performance. Therefore, further post-processing of these candidate sequences is crucial.}  Specifically,
if two mappings, $\tau^{'}_{1}$ and $\tau^{'}_{2}$ (more mappings of the same value are also applicable), have the same $Gval$ value, specifically $Gval(\tau^{current},{seq}_{{opt}} ^1)= Gval (\tau^{current},{seq}_{opt}^2)$. Then, an evaluation function based on weight and distance is proposed to assess their merits, thereby improving the routing algorithm. Given that gates in the rear have a relatively small impact on the mapping algorithm, and considering the potentially large number of subsequent gates, a significant amount of additional calculations will be introduced. Therefore,  a look-ahead window is set, with its size determined by the hyperparameter $wnd$.  {The post-processing evaluation function is defined as follows, with the objective of maximizing its value:} 
\begin{equation}\label{eq5}
\begin{split}
% &argmax_{\tau^{'} } \ dw(\tau^{'} )=\sum_{g_i\ in\ LC}^{window\_size}W(g_i )* \\&(diameter(G) -dis(\tau^{'}(q),\tau^{'}(p))),\\&
 dw(\tau^{'})=\sum_{g_i\, \in \, LC}^{wnd}W_g(g_i )*(dia(AG)-dis(\tau^{'}(q),\tau^{'}(p))),
\end{split} 
\end{equation}
where $W_g(g_i)$ represents the weight of gate $g_i$, $dis(\tau^{'}(q),\tau^{'}(p))$ represents the distance of qubit $q$ and $p$ in the gate $g_i$ under the mapping relationship $\tau^{'}$.
% The design of this evaluation function can comprehensively 
\begin{figure*}[t]
\centering
\includegraphics[width=0.9\linewidth]{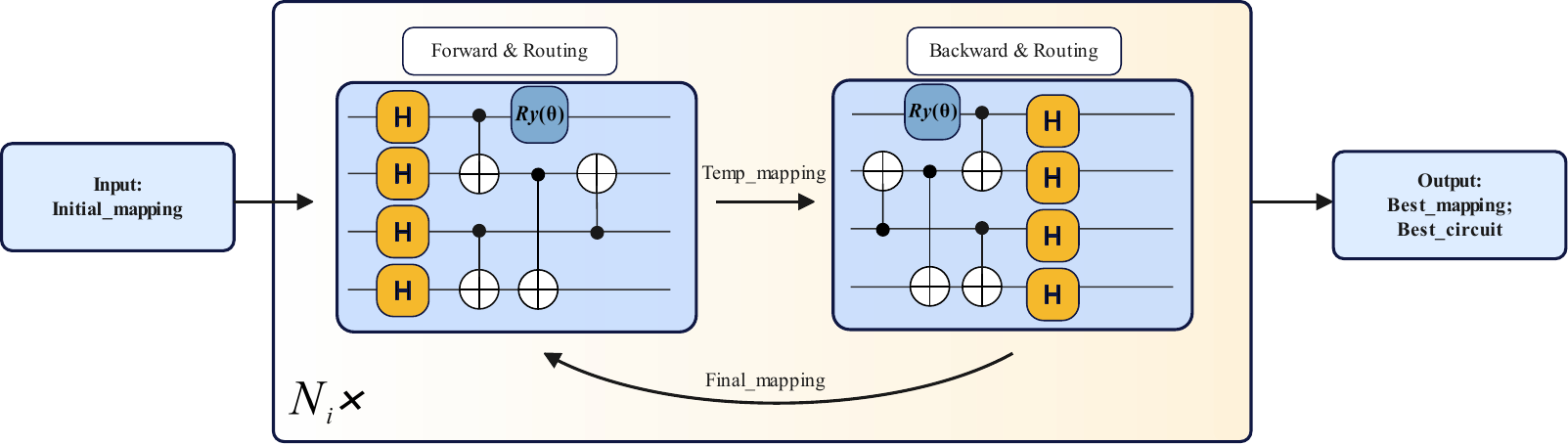}
\caption{A simplified iterative optimization framework for the whole qubit mapping problem.}
\label{fig_5}
\end{figure*}

 {The motivation behind designing this function is to prioritize the execution of quantum gates located at the front of the circuit during the mapping process, thereby facilitating the execution of subsequent gates. This is the main reason that we focus on high-priority quantum gates and strive to minimize the distance between their corresponding qubits after the mapping transformation. Through the effective integration of these two factors, the proposed function enables further optimization of SWAP gate sequence selection by maximizing its value.} Algorithm. \ref{alg3} combines the SWAP sequence selection and post-processing to optimize intermediate qubit routing. It takes inputs such as the logical circuit, the physical hardware architecture, and the generated initial mapping. Initially, it identifies executable gates (EG) based on the current mapping, transferring them from the logical circuit $LC$ to the physical circuit $PC$. The algorithm iterates until all gates in $LC$ are executed. During each iteration, it calculates the SWAP sequence with the largest $Gval$ and applies post-processing to find the most suitable sequence using the discriminant function $dw$. The selected SWAP gates are then inserted into $PC$ sequentially, and the current mapping is adjusted according to the SWAP sequence using the transformation function. This process continues until all gates are executed and the modified physical circuits $PC$ and $\tau^{current}$ are returned.  

To clearly illustrate the entire SWAP sequence insertion process, the schematic diagram of the insertion of the SWAP sequence is shown in Figure \ref{fig_8}. Under the mapping $\tau^{current}$, two CNOT gates can be executed, while four CNOT gates (marked in red) cannot. We choose a Depth-2 sequence search in this example. After applying the heuristic evaluation functions in Eq. (\ref{eq4}) and Eq. (\ref{eq5}), the optimal SWAP sequence is selected and inserted into the circuit, resulting in the mapping \( \tau' \), which enables the execution of the subsequent three CNOT gates.

In general, the intermediate qubit routing algorithm can take into account many factors such as the size of the search space, weight information, and distance between qubits. Compared with many other intermediate qubit routing algorithms, it can comprehensively assess the influence of {weight and distance} and effectively balance their relationships. This also enables the mapping algorithm to outperform many existing algorithms in terms of overall effectiveness.

\subsection{Simplified Iterative Optimization Framework
}
 {So far, the initial mapping and intermediate qubit routing yield a physical circuit that meets the device’s connectivity constraints. However, because these methods are heuristic, the solution related to SWAP cost may not be optimal. To efficiently approach a near-optimal solution without incurring substantial computational cost, we propose a simplified iterative optimization framework, as shown in Figure \ref{fig_5}. This framework is based on the common forward-backward traversal technique \cite{li2019tackling}. Specifically, after a well-modified initial mapping is given, the routing algorithm first operates on the original circuit (OC), referred to as forward traversal, producing a $Temp\_mapping$ that naturally satisfies the connectivity constraints of the quantum gates at the end of the OC. Subsequently, the $Temp\_mapping$ is input into the reverse circuit of OC for routing, which we refer to as backward traversal. The resulting $Final\_mapping$ is then used as a new initial mapping for routing on the OC, completing one iteration of the optimization process.} Ultimately, our objective is to preserve the best mapping and quantum circuit with the smallest number of additional SWAP gates inserted throughout the iterations. Within the iterative framework, the parameter $N_i$ indicates the number of iterations required for optimization. Importantly, it is crucial to track the circuit execution direction at minimal cost during the iteration process.  This is because the case with the minimum number of SWAP gates inserted may occur in the backward-traverse phase, where it is necessary to invert the current circuit to obtain the target quantum circuit.

Compared to the iterative algorithm presented in \cite{zhu2021iterated},  {our framework incorporates an effective initial mapping instead of a random one, ensuring that the search space of initial mapping is closer to the set of reasonable solutions. This accelerates convergence to the optimal mapping, enabling the algorithm to achieve superior results with fewer iterations compared to previous iterative frameworks. Moreover, since the routing algorithm largely determines the upper bound of the mapping algorithm's performance, it is wise to allocate more computational resources to each routing step. This strategic allocation allows the algorithm to rapidly converge to a near-optimal solution within a limited number of iterations Given the inherently low iteration requirement of the framework, we further simplify certain mapping shuffling operations after the routing process, as excessive shuffling can lead the mapping to deviate from the predefined reasonable solution set, ultimately reducing efficiency and performance. Consequently, our framework requires only a few circuit traversals using the routing algorithm to identify the optimal solution, whereas previous iterative frameworks often involve extensive traversals, resulting in unnecessary computational overhead.}

\textbf{Complexity Analysis.} As for the analysis of the algorithm's time complexity. In the worst-case scenario, for any given architecture $AG(V, E)$ and quantum circuit $LC(Q, G)$, if the SWAP sequences have length $l$, there will be $O(|E|^l)$ possible sequences. The computational complexity of each sequence is linearly related to $wnd$, and if only one gate is executed each time, there will be a maximum of $|G|$ iterations. The entire process requires $N_i$ iterations. Therefore, the overall time complexity of the iterative framework, including routing, is $O(N_i \cdot |E|^l \cdot wnd \cdot |G|)$.

\section{Experiments and Evaluation}
In this section, we will evaluate the proposed HAIL algorithm and compare its performance with that of other algorithms across various benchmarks and coupling architectures.

\textbf{Benchmarks.} We evaluate the proposed algorithm using publicly available benchmarks, including $\mathcal{B}_{23}$, which contains 23 quantum circuits categorized by gate count: small ( $<300$ gates), medium ($300-4000$ gates), and large ($\ge4000$ gates). Additionally, we assess the algorithm's performance across various scenarios using multiple benchmarks \cite{zhou2020monte}. These benchmarks, limited to 20 qubits, include single-qubit operations (e.g., $H$, $T$, $Ry(\theta)$) and CNOT gates. A detailed description of these benchmarks is provided in Table \ref{tab:benchmark}, allowing for a comprehensive evaluation of the algorithm across different circuit scales.
\begin{table}[ht]
\caption{Characteristics of the benchmark set}
    \centering
    \scalebox{0.8}{ % Resize to column width
        \begin{tabular}{m{2.5cm} c c} % Set fixed column widths
            \toprule
   {\centering Benchmark Name
    /Number of Circuits}& {Total Number of Gates}
            & {Average CNOTs per Circuit} \\ \midrule
            \centering $\mathcal{B}_{23}$/23        & 63333    & 1202.609   \\
            \centering $\mathcal{B}_{114}$/114      & 554497   & 2180.289   \\
            \centering $\mathcal{B}_{ran}$/170      & 34000    & 200        \\
            \centering $\mathcal{B}_{real}$/173     & 603654   & 1506.295   \\ \bottomrule
        \end{tabular}}
    \label{tab:benchmark}
\end{table}

\textbf{Hardware Architecture.}
Commonly used quantum architectures, such as the IBM Q20 (Figure \ref{fig_4}(c)) and the Google Sycamore (Figure \ref{fig_4}(d)), were employed to evaluate the algorithm’s adaptability by considering architectures with varying levels of sparsity. 

\textbf{Algorithm Configuration.} To distinguish between different SWAP search depths during the routing phase, we defer to the algorithm with a search depth of $n$ as HAIL-$n$. The number of iterations $N_i$ is fixed at 5, and each SWAP gate involves at least one qubit from the first three layers of the remaining gates. Additionally, to adjust $wnd$ dynamically, if the remaining gates $rg$ exceed 4000, $wnd$ is set to $\lfloor 1.5*\sqrt{rg}\rfloor$, otherwise, it remains at 30.

\textbf{Experimental Platform.}
All experiments were conducted on a personal laptop featuring an Intel(R) Core(TM) i5-8300H CPU @ 2.30GHz, paired with 16 GB of DDR4 RAM.

\textbf{Comparative Algorithms.}
To validate the effectiveness of the proposed algorithm, we compare it with state-of-the-art algorithms like SABRE\cite{li2019tackling}, ILS\cite{zhu2021iterated}, TWP\cite{qian2023method}. Since TWP is an improvement of FIDLS\cite{li2020qubit} and exhibits superior performance, it has been selected for comparison.

\makeatletter
\newsavebox{\@tabnotebox}
\providecommand\tmark{} % so having ctable or not is irrelevant
\providecommand\tnote{}
\newenvironment{tabularwithnotes}[3][c]
{\long\def\@tabnotes{#3}%
\renewcommand\tmark[1][a]{\makebox[0pt][l]{\textsuperscript{\itshape##1}}}%
\renewcommand\tnote[2][a]{\textsuperscript{\itshape##1}\,##2\par}
\begin{lrbox}{\@tabnotebox}
\begin{tabular}{#2}}
{\end{tabular}\end{lrbox}%
\parbox{\wd\@tabnotebox}{
\usebox{\@tabnotebox}\par
\smallskip\@tabnotes
}%
}
\makeatother
\begin{table*}
\renewcommand{\arraystretch}{1.4}
\label{table1}
\caption{Number of additional gates compared with SABRE\cite{li2019tackling}, ILS\cite{zhu2021iterated}, and TWP\cite{qian2023method} algorithms on benchmark $\mathcal{B}_{23}$.}

\centering

\renewcommand{\arraystretch}{1.2}
\begin{tabular}{cccccccccc}

% {\linewidth}{@{}LLLLLLLLLL@{}}

\toprule
\multicolumn{3}{c}{Benchmark}&
 \multicolumn{4}{c}{Number of additional CNOT gates} &  \multicolumn{3}{c}{Optimization ratio} \\ 
 Circuit Name&	Qubit Num&	Gate Num&	 $g_1$\cite{li2019tackling}&	$g_2$\cite{zhu2021iterated}&	$g_3$\cite{qian2023method}&$g_0$(HAIL-3)&	$(g_1-g_0)/g_1$ &	$(g_2-g_0)/g_2$& $(g_3-g_0)/g_3$
									
\\
\midrule
        4mod5\_v1\_22 & 5 & 21 & 0 & 0 & 0 & 0 & 0.00\% & 0.00\% & 0.00\% \\ 
        mod5mils\_65 & 5 & 35 & 0 & 0 & 0 & 0 & 0.00\% & 0.00\% & 0.00\% \\ 
        alu-v0\_27 & 5 & 36 & 3 & 3 & 9 & 3 & 0.00\% & 0.00\% & 66.67\% \\
        decod24v2\_43 & 4 & 52 & 0 & 0 & 0 & 0 & 0.00\% & 0.00\% & 0.00\% \\ 
        4gt13\_92 & 5 & 66 & 0 & 0 & 0 & 0 & 0.00\% & 0.00\% & 0.00\% \\ 
        qft\_10 & 10 & 200 & 54 & 27 & 33 & 30 & 44.44\% & -11.11\% & 9.09\% \\ 
        qft\_16 & 16 & 512 & 186 & 120 & 117 & 120 & 35.48\% & 0.00\% & -2.56\% \\ 
        ising\_model\_10 & 10 & 480 & 0 & 0 & 0 & 0 & 0.00\% & 0.00\% & 0.00\% \\ 
        ising\_model\_13 & 13 & 633 & 0 & 0 & 0 & 0 & 0.00\% & 0.00\% & 0.00\% \\ 
        ising\_model\_16 & 16 & 786 & 0 & 0 & 0 & 0 & 0.00\% & 0.00\% & 0.00\% \\ 
        rd84\_142 & 15 & 343 & 105 & 60 & 66 & 57 & 45.71\% & 5.00\% & 13.64\% \\ 
        sym6\_145 & 7 & 3888 & 1272 & 495 & 405 & 369 & 70.99\% & 25.45\% & 8.89\% \\
        z4\_268 & 11 & 3073 & 1365 & 456 & 363 & 372 & 72.75\% & 18.42\% & -2.48\% \\ 
        radd\_250 & 13 & 3213 & 1275 & 579 & 576 & 384 & 69.88\% & 33.68\% & 33.33\% \\ 
        cycle10\_2\_110 & 12 & 6050 & 2622 & 948 & 969 & 729 & 72.20\% & 23.10\% & 24.77\% \\ 
        adr4\_197 & 13 & 3439 & 1614 & 516 & 648 & 381 & 76.39\% & 26.16\% & 41.20\% \\ 
        misex1\_241 & 15 & 4813 & 1521 & 492 & 444 & 414 & 72.78\% & 15.85\% & 6.76\% \\ 
        rd73\_252 & 10 & 5321 & 2133 & 876 & 732 & 615 & 71.17\% & 29.79\% & 15.98\% \\ 
        square\_root\_7 & 15 & 7630 & 2598 & 819 & 1017 & 738 & 71.59\% & 9.89\% & 27.43\% \\
        co14\_215 & 15 & 17936 & 8982 & 2775 & 2658 & 2340 & 73.95\% & 15.68\% & 11.96\% \\ 
        rd84\_253 & 12 & 13658 & 6147 & 2484 & 1827 & 1719 & 72.04\% & 30.80\% & 5.91\% \\ 
        sqn\_258 & 10 & 10223 & 4344 & 1599 & 1212 & 999 & 77.00\% & 37.52\% & 17.57\% \\ 
        sym9\_193 & 11 & 34881 & 16653 & 5562 & 5361 & 3777 & 77.32\% & 32.09\% & 29.55\% \\ 
        Sum & - & - & \bf{50874} & \bf{17811} & \bf{16437} & \bf{13047} & \bf{74.35}\% & \bf{26.75}\% & \bf{20.62}\% \\ 
\bottomrule
\end{tabular}

\end{table*}

\subsection{Performance Analysis Compared with Some Algorithms on IBM Q20}
This experiment was conducted on the IBM Q20 architecture using the $\mathcal{B}_{23}$ benchmark. We can see in Table \ref{table1}, the first three columns contain the circuit benchmark information, while the subsequent four columns ($g_1$, $g_2$, $g_3$, $g_0$) respectively display the number of CNOT gates inserted by SABRE, ILS, TWP, and our proposed algorithm HAIL-3 into the corresponding circuit. The last three columns show the improvement rate of our algorithm to others. This comparison is calculated by the formula:
\begin{equation}
\label{eq6}
\Delta_i=1- \frac{g_0}{g_i}.
\end{equation}

From the comparative data, it is evident that the optimization capability of our proposed HAIL-3 algorithm is limited for small-scale circuits, as the competing algorithms are also able to find optimal solutions in these cases. Specifically, when compared to the TWP algorithm, our algorithm achieves a 66.67\% improvement in the number of additional CNOT gates on the quantum circuit $alu-v0\_27$. However, for the $qft\_10$ circuit, when compared to the ILS algorithm, our algorithm results in a negative optimization effect, which may be due to the choice of the look-ahead window size. Given the small scale of the quantum circuits and the fact that the difference in the number of additional CNOT gates is minimal, such an optimization rate is still considered acceptable.

Conversely, for medium-scale or large-scale circuits, Compared to the SABRE, ILS, and TWP algorithms, our proposed algorithm demonstrates notable optimization improvements on these two types of quantum circuits.
Specifically, HAIL-3 achieves improvements of 77.32\%, 32.09\%, and 29.55\%, respectively, over the three algorithms in the $sym9\_193$ quantum circuit, representing a significant enhancement. Moreover, this algorithm outperforms SABRE and ILS in all corresponding circuits, and similarly, it performs better than the TWP algorithm in most cases, with the exception of a slight performance decrease observed in $z4\_268$. The comparison with TWP shows a -2.48\% optimization rate, which highlights the limitations of heuristic algorithms, as they are prone to getting trapped in local optima, leading to unstable optimization results.

According to the data in the table, the proposed algorithm has a 74.35\% performance improvement compared to the foundational SABRE algorithm. Similarly, for ILS, which is also an iterative algorithm, the performance improves by 26.75\%. Additionally, TWP, the recently proposed algorithm with the best comprehensive effect, also has an optimization of 20.62\%. Therefore, the algorithm in this paper has a corresponding performance improvement on most of the quantum circuits in the benchmark, and this improvement is more obvious in medium and large quantum circuits.

\subsection{Trade-Off Between Performance and Time}

In previous experiments, it was observed that when the search depth exceeded 3 in HAIL, the algorithm's runtime increased significantly, which is attributed to the exponential dependence on the search depth. However, we found that a search depth of 3 typically provides a balance between performance and runtime, making it generally acceptable. Yet, for larger circuits, this setting still resulted in longer runtime compared to other heuristic algorithms. Through multiple experiments, it was noticed that SWAP sequences with a depth of 3 occurred relatively infrequently throughout the decision-making process. On the other hand, considering only the Depth-2 SWAP sequences led to a significant decline in performance. 

 To further enhance the scalability of the algorithm and achieve a better balance between performance and runtime, we propose a partially extended strategy to narrow the candidate set of SWAP sequences, referring to it as HAIL-imp.  {The core idea behind this approach is to establish connections between low-depth and high-depth sequences through an extension process, rather than treating sequences of different depths as independent samples and sampling them separately. This strategy enables more effective integration of SWAP sequences across varying depths, thereby enhancing the decision-making process for selecting the optimal SWAP sequences.}
\begin{figure}[!t]
\centering
\includegraphics[width=\linewidth]{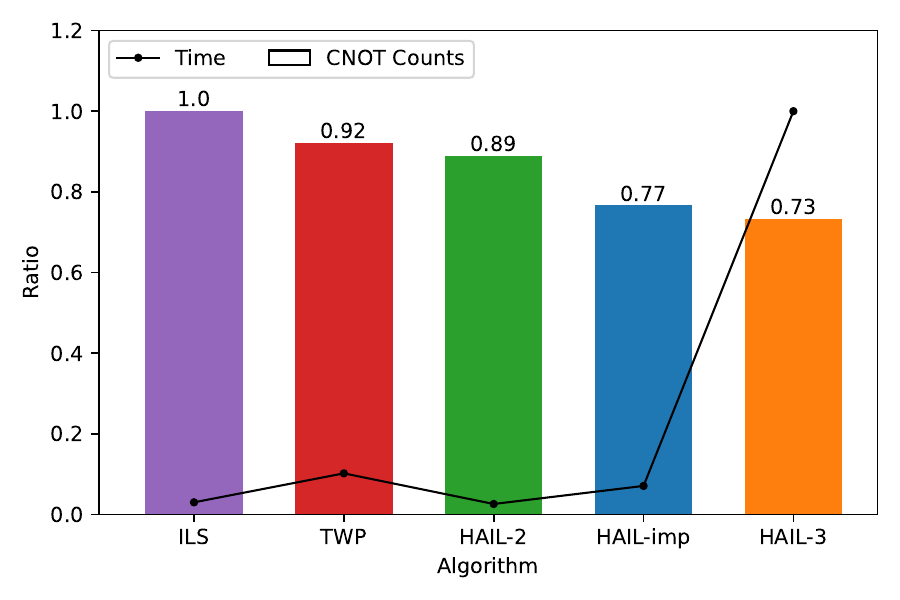}
\caption{{Comparison of the proposed HAIL series algorithms, TWP, and ILS on the IBM Q20 for $\mathcal{B}_{23}$, with the inserted CNOT gates normalized to ILS and the algorithm runtime normalized to HAIL-3. }
}

\label{fig_b23_q20}
\end{figure}
{Specifically, We start with a Depth-2 sequence search and select the top $K$ sequences based on Eq.\ref{eq4} for evaluation. Among these $K$ sequences, only those with a length of 2 will be extended to a length of 3, as they demonstrate greater potential. In the worst case, the number of sequences of length 3 is $K * |EG|$, and the total number of all sequences is $O(|EG|^2 + (K+1) * |EG|)$. This is significantly smaller than the number of sequences generated with a maximum search depth of 3, which is $O(|EG|^3)$. In the experiments, $K$ is set to 50. All these algorithms are tested on the same benchmark, $\mathcal{B}_{23}$, using the IBM Q20 architecture. As shown in Figure \ref{fig_b23_q20}, the improved search algorithm HAIL-imp significantly reduces time with only a slight decrease in overall performance when compared to HAIL-3. Besides, our proposed HAIL outperforms both TWP and ILS in additional inserted SWAP gates when the maximum search depth is 2, and as the search depth increases, the runtime also increases. Specifically, while HAIL-imp results in a 5\% increase in CNOT gate insertions compared to HAIL-3, it achieves a 90\% reduction in execution time, which represents the desired balance between performance and runtime. Similarly, both the HAIL-2 and ILS algorithms exhibit a slight decrease in execution time compared to HAIL-imp, but they lead to increases in CNOT gate insertions by 15.58\% and 29.87\%, respectively. In comparison to the TWP algorithm, HAIL-imp not only reduces execution time but also decreases CNOT gate insertions by 16.3\%.}

\begin{figure}[!t]
\centering
\hspace*{-0.5cm}
\includegraphics[width=1.1\linewidth]{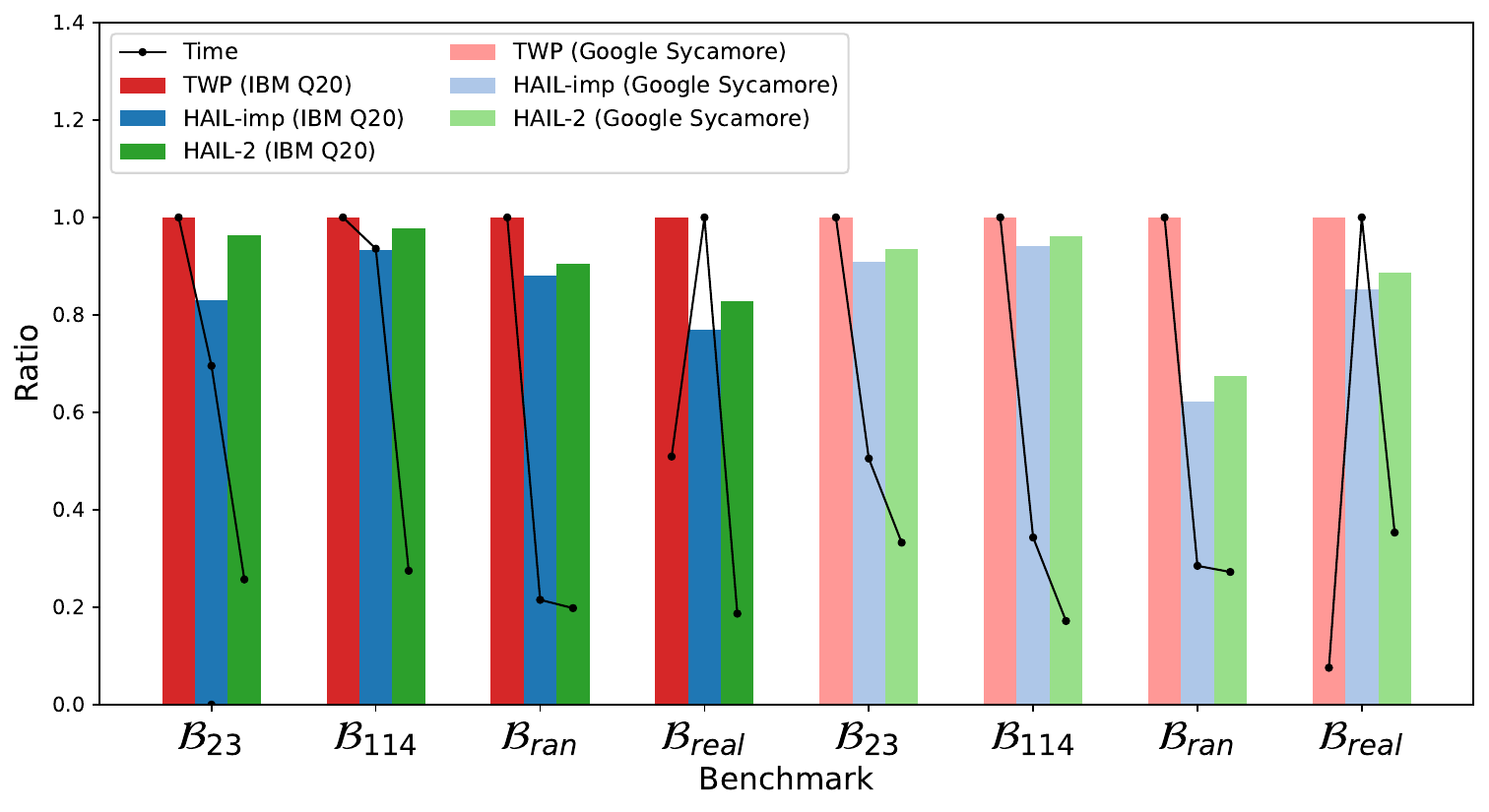}
\caption{Comparison of the proposed HAIL-X series methods with TWP on the benchmarks $\mathcal{B}_{23}$, $\mathcal{B}_{114}$, $\mathcal{B}_{{ran}}$, and $\mathcal{B}_{{real}}$ is conducted on the IBM Q20 and Google Sycamore architectures, with normalization of the results based on the number of inserted CNOT gates(bar chart) and algorithm runtime(line chart).
}
\label{fig_q20&sycamore}
\end{figure}

{Additionally, to demonstrate the feasibility of the algorithm across different benchmarks, we conduct experiments on the $\mathcal{B}_{114}$, $\mathcal{B}_{ran}$, and $\mathcal{B}_{real}$. Since the algorithm TWP outperforms both ILS and SABRE, it is compared only with the proposed algorithm HAIL in these benchmarks, with experiments performed on the IBM Q20 and Sycamore architectures. The experimental results on IBM Q20 are shown in the Figure \ref{fig_q20&sycamore}. According to the results, the HAIL-imp algorithm achieved an approximate 8\% reduction in both runtime and the number of additional CNOT gates on the $\mathcal{B}_{114}$ benchmark. On $\mathcal{B}_{ran}$, the number of CNOT gates decreases by 12\%, while the runtime is significantly reduced by 78.5\%, marking a considerable improvement. However, on the $\mathcal{B}_{real}$, there is a 23.07\% reduction in additional CNOT gates, but the required runtime increases. This may be due to the characteristics of the benchmark, leading to fewer gate executions per decision-making, which in turn increases the whole time. This issue may need to be addressed in future work. Overall, on the IBM Q20 for various benchmarks, HAIL-imp shows a reduction in the number of CNOT gate insertions and time, which also reflects the effectiveness of the algorithm. As for HAIL-2, the algorithm consistently demonstrates fewer CNOT gates and shorter runtime compared to the TWP in these scenarios, but the additional number of CNOT gates inserted is higher than that of HAIL-imp.}

Similarly, the experimental results on Sycamore are shown in Figure \ref{fig_q20&sycamore}. {As we can see both the HAIL-imp and HAIL-2 algorithms exhibit a reduction in the number of additional CNOT gates inserted compared to the TWP algorithm for the four specified benchmarks on the Sycamore architecture. This reduction is more obvious in the $\mathcal{B}_{ran}$ benchmark, where HAIL-imp and HAIL-2 achieve reductions of $37.8\%$ and $32.6\%$, respectively. For the remaining benchmarks, a reduction of approximately 10\% is also observed. Notably, the HAIL-imp algorithm consistently inserts fewer additional SWAP gates than HAIL-2. In terms of runtime, the HAIL-imp algorithm demonstrates a longer execution time solely on the $\mathcal{B}_{real}$ benchmark compared to both HAIL-2 and TWP. Besides, it is reasonable that HAIL-imp has a longer runtime compared to HAIL-2, as the algorithm's search space is larger, evaluating a broader range of candidate SWAP sequences. For the other three benchmarks, the runtime of the HAIL-imp algorithm is at least $50\%$ shorter than that of the TWP algorithm, further highlighting the effectiveness and scalability of the proposed algorithms. }

\section{Conclusion}
 {In this paper, we propose HAIL, an iterative heuristic algorithm designed to minimize the insertion of additional SWAP gates to tackle the qubit mapping problem, which can be divided into initial mapping and intermediate qubit routing. Leveraging the parallelism of quantum gates within the circuit, we propose a layer-weight assignment for each quantum gate at different layer accordingly, which is then integrated with a subgraph isomorphism algorithm to construct the initial mapping. Since the subgraph isomorphism may result in partially unmapped qubits, a mapping completion algorithm is introduced to address this issue by effectively balancing gate weights and qubit distance properties. For the qubit routing stage, a SWAP gate sequence search is introduced that considers the average number of executable CNOT gates per inserted SWAP gate to identify promising SWAP candidates. Building upon this, we propose a novel scoring function that also accounts for gate weights and qubit distance properties, and incorporates a variable look-ahead window, adapted to the number of remaining gates in the circuit, to guide the identification of the optimal SWAP sequence and reduce the insertion of SWAP gates. Additionally, to address the suboptimality of a single run of the routing algorithm, we introduce a simplified iterative framework that leverages forward–backward traversals with a few iterations to identify a more optimal SWAP gate insertion situation. The performance of the HAIL algorithm is evaluated through a comparative analysis with state-of-the-art algorithms, including SABRE, ILS, and TWP, on the IBM Q20 architecture. Experiments on the benchmark $\mathcal{B}_{23}$ demonstrate that HAIL-3 reduces the number of additional SWAP gates by 20.62\% compared to TWP and also outperforms SABRE and ILS.}

 {To further optimize the trade-off between runtime and performance in HAIL, we propose an improved variant named HAIL-imp. Building upon the HAIL-2 algorithm, HAIL-imp selectively extends certain length-two sequences to length-three, given that the probability of selecting length-three sequences is relatively low. HAIL-imp reduces the search space of candidate SWAP sequences, thereby improving search efficiency compared to HAIL-3. Evaluation on both the IBM Q20 and Google Sycamore quantum processors demonstrates significant improvements in runtime and the number of additional SWAP gates compared to TWP. On the $\mathcal{B}_{ran}$ benchmark under the Sycamore architecture, the number of additional SWAP gates was reduced by 37.8\%, and runtime decreased by 71.52\%, confirming the effectiveness and scalability of the algorithm.
}

\section*{acknowledgments}
This research was supported by the National Nature Science Foundation of China (Grant No. 62101600, Grant No. 62471070), the Science Foundation of the China University of Petroleum, Beijing (Grant No. 2462021YJRC008), and the State Key Lab of Processors, Institute of Computing Technology, CAS (Grant No. CLQ202404).

% \clearpage
% \bibliographystyle{IEEEtran}
% \bibliography{IEEEabrv,ref}

\bibliographystyle{IEEEtran}

\end{CJK}
\end{document}